\documentclass[superscriptaddress,showpacs,preprint,preprintnumbers,amsmath,amssymb]{revtex4}

\usepackage{graphicx}
\usepackage{dcolumn}
\usepackage{bm}
\usepackage{amsmath}
\usepackage{epsfig}
\usepackage{rotating}
\usepackage{enumerate}
\usepackage{xcolor}

\begin{document}

\title{Giant electrophononic response in PbTiO$_3$ by strain engineering}

\author{Pol Torres}
\affiliation{Institut de Ci\`encia de Materials de Barcelona (ICMAB--CSIC)
             Campus de Bellaterra, 08193 Bellaterra, Barcelona, Spain}

\author{Jorge \'I\~{n}iguez}
\affiliation{Materials Research and Technology Department, Luxembourg
             Institute of Science and Technology (LIST),
             Avenue des Hauts-Fourneaux 5, L-4362 Esch/Alzette, Luxembourg}
\affiliation{Physics and Materials Science Research Unit, University of Luxembourg,
              41 Rue du Brill, L-4422 Belvaux, Luxembourg}

\author{Riccardo Rurali}
\affiliation{Institut de Ci\`encia de Materials de Barcelona (ICMAB--CSIC)
             Campus de Bellaterra, 08193 Bellaterra, Barcelona, Spain}
\email{rrurali@icmab.es}

\date{\today}

\begin{abstract}
We demonstrate theoretically how, by imposing epitaxial strain in 
a ferroelectric perovskite, it is possible to achieve a dynamical 
control of phonon propagation by means of external electric fields, 
which yields a giant electrophononic response, i.e. the dependence 
of the lattice thermal conductivity on external electric fields. 
Specifically, we study the strain-induced manipulation of the lattice 
structure and analyze its interplay with the electrophononic response.
We show that tensile biaxial strain can drive the system to a regime
where the electrical polarization can be effortlessly rotated and
thus yield giant electrophononic responses that are at least one order of magnitude 
larger than in the unstrained system. These results
derive directly from the almost divergent behavior of the electrical
susceptibility at those critical strains that drive the polarization
on the verge of a spontaneous rotation.
\end{abstract}

\maketitle

Heat in insulators and semiconductors is carried by phonons,
the quanta of lattice vibrations,
and the thermal conductivity is determined by the associated
dissipative processes.
The manipulation of phonons and the dynamical tuning of the thermal
conductivity of a solid are problems of fundamental interest in
condensed matter physics~\cite{LiRMP12,VolzEPJB16,ZardoCOGSC19} and 
have important implications in renewable energy applications --such 
as vibrational energy harvesting~\cite{PopNR10},  
thermoelectricity~\cite{BenentiPR17} or electrocaloric cooling~\cite{MoyaMRS18}-- and for the implementation 
of a phonon-based logic, which relies on thermal diodes~\cite{TerraneoPRL02,
LiPRL04} and transistors~\cite{LiAPL06}, and where information is 
transmitted and processed by heat carriers.

Ferroelectric materials favor a spontaneous lattice distortion, 
below a critical temperature, which has an associated dipole moment 
that can be controlled with an external electric field. Therefore, 
they are the ideal playground to explore phonon manipulation, because the 
modifications of the lattice structure translate directly into changes 
of the vibrational properties and thus of the thermal conductivity. 
The polarization, ${\mathbf P}$, can be selectively oriented, for instance, creating 
neighboring regions separated by domain walls, which may act as phonon 
scatterers or filters~\cite{SeijasBellidoPRB17,RoyoPRM17}. More generally, 
an electric field can {\it strengthen} or {\it weaken} ${\mathbf P}$, when 
it is parallel to it, or partially rotate it, when it has a component 
perpendicular to it~\cite{SeijasBellidoPRB18,TorresPRM19,SeijasBellido19}. 
This {\it electrophononic} effect, whereby an electric field is used to 
tune the thermal conductivity via a controlled modification of the 
crystal lattice, paves the way toward an all-electrical control of 
the heat flux. 

The temperature- and field-dependent thermal conductivity, 
$\bm{\kappa}$, can be written as a second-order expansion 
in terms of the thermal-response tensors $\bm{\alpha}$ and 
$\bm{\beta}$ as
\begin{equation}
\kappa_{ij}(T,{\mathbf E}) = \kappa_{ij}^{0}(T) + \sum_{k}
\alpha_{ij,k}(T) E_{k} + \sum_{kl} \beta_{ij,kl}(T) E_{k}E_{l} \; ,
\label{eq:response}
\end{equation}
where $i$, $j$, $k$, and $l$ are the spatial directions $x$, $y$, 
and $z$ and $\bm{\kappa}^{0}$ is the conductivity at zero 
applied field. The physical mechanisms that lead to a coupling between
${\mathbf E}$ and $\bm{\kappa}$ are different for fields parallel
or perpendicular to ${\mathbf P}$. As shown in Ref.~\onlinecite{SeijasBellidoPRB18}
by some of us, in the former case the applied
field results in a hardening of the phonon frequencies throughout
the whole spectrum, if ${\mathbf E}$ is parallel to  
${\mathbf P}$, or a softening, if it is anti-parallel to it.
In the latter case the main effect of the applied field is lowering
the symmetry of the lattice, thus leading to a larger phase-space 
for phonon-phonon anharmonic scattering processes~\cite{SeijasBellidoPRB18,
TorresPRM19}. The reported changes in the thermal conductivity are large,
but typically require applied electric fields that are substantial.
In this Letter we study the dependence of the electrophononic
coefficients $\bm{\alpha}$ and $\bm{\beta}$ on epitaxial strain and show
that suitable strain conditions can increase them by one order of
magnitude and result, in principle, in nearly divergent responses at certain critical strains,
where a vanishing small electric field leads to large variation
of the thermal conductivity.

We focus on PbTiO$_3$ (PTO), a paradigmatic ferroelectric oxide 
with a tetragonal perovskite structure and a critical temperature 
of 760~K, and consider biaxial tensile strains, $\epsilon$, in the 
plane perpendicular to the tetragonal axis of the ground state up to 3$\%$. Note that biaxial
strains within this range can be routinely achieved by epitaxially growing
PTO thin films on appropriate substrates~\cite{SchlomARMR07}; further, the strain
can also be controlled dynamically if a piezoelectric substrate is 
used~\cite{IidaAPE19,ZhengPRB07}.

We calculate the ground-state structure, the harmonic and third-order 
anharmonic force constants (IFCs) within second-principles density-functional 
theory (SPDFT). The term second-principles refer to methods that are 
first-principles-based both in their formulation and in the way the 
information needed to use them is obtained~\cite{GarciaFernandezPRB16}. 
We use the implementation of the SCALE-UP code~\cite{WojdelJPCM13,
GarciaFernandezPRB16}, which relies on polynomial potentials fitted 
from DFT calculations.
As most first-principles approaches, SPDFT 
reproduces accurately the vibrational and response properties of 
PTO~\cite{WojdelJPCM13} and it has a documented predictive 
power for the most important structural, vibrational and response 
properties of ferroelectric perovskite oxides~\cite{ZubkoNature16,
ShaferPNAS18}. We compute the IFCs in supercells defined as an $8 \times 8 \times 8$ 
repetition of the 5-atom unit cell using the finite differences method. 
For the harmonic displacements we use the {\sc Phonopy} code~\cite{TogoSM15} 
considering all neighboring interactions. {\sc thirdorder.py}~\cite{LiPRB12b,LiCPC14} is used 
to determine the anharmonic interactions, neglecting those beyond 
twelfth neighbors, a choice that we checked yields 
converged results. The IFCs calculated within the SPDFT framework 
are used to obtain an iterative solution of linearized phonon 
Boltzmann Transport Equation (BTE)~\cite{WardPRB09} with the {\sc ShengBTE} 
code~\cite{LiCPC14}, thus avoiding the shortcomings of the relaxation 
time approximation (RTA) that treat Normal processes as resistive 
scattering events~\cite{WardPRB09,Torres2DMater19}. The thermal conductivity is then expressed as:
\begin{equation}\label{eq_iterative}
\kappa_{ij}=\frac{1}{k_b T^2 N \Omega}\sum_{\lambda}n_{\lambda}(n_{\lambda}+1)(\hbar \omega_{\lambda})^2 v_{i,\lambda}\tau_{\lambda}(v_{j,{\lambda}}+\Delta_{j,\lambda}) \;,
\end{equation}

where $N$ is the number of $\textbf{q}$-points used in the sampling
of the Brillouin zone, $\Omega$ the volume of the 5-atom perovskite
unit cell.
The sum runs over all phonon modes, the index
$\lambda$ including both ${\bf q}$-point and phonon band.
$n_{\lambda}$ is the Bose-Einstein 
distribution function, and $\omega_{\lambda}$, $\tau_{\lambda}$ 
and $v_{i,\lambda}$ correspond to the phonon frequency, relaxation 
time and group velocity, $\partial \omega_\lambda / \partial q_i$,
respectively. The term $\Delta_{j,\lambda}$ 
takes into account the deviation of the heat current with respect 
to the RTA approach and is thus relevant on those systems where Normal
processes play an important role. We solve Eq.~(\ref{eq_iterative}) 
on a $8 \times 8 \times 8$ {\bf q}-point grid and obtain the thermal
conductivity, $\bm{\kappa}(T)$, by summing over all the modes.
We also include isotopic disorder scattering, using the natural
abundances of isotopes of Pb, Ti, and O, within the model of 
Tamura~\cite{TamuraPRB83}.

\begin{figure}[t]
\includegraphics[width=0.7\linewidth]{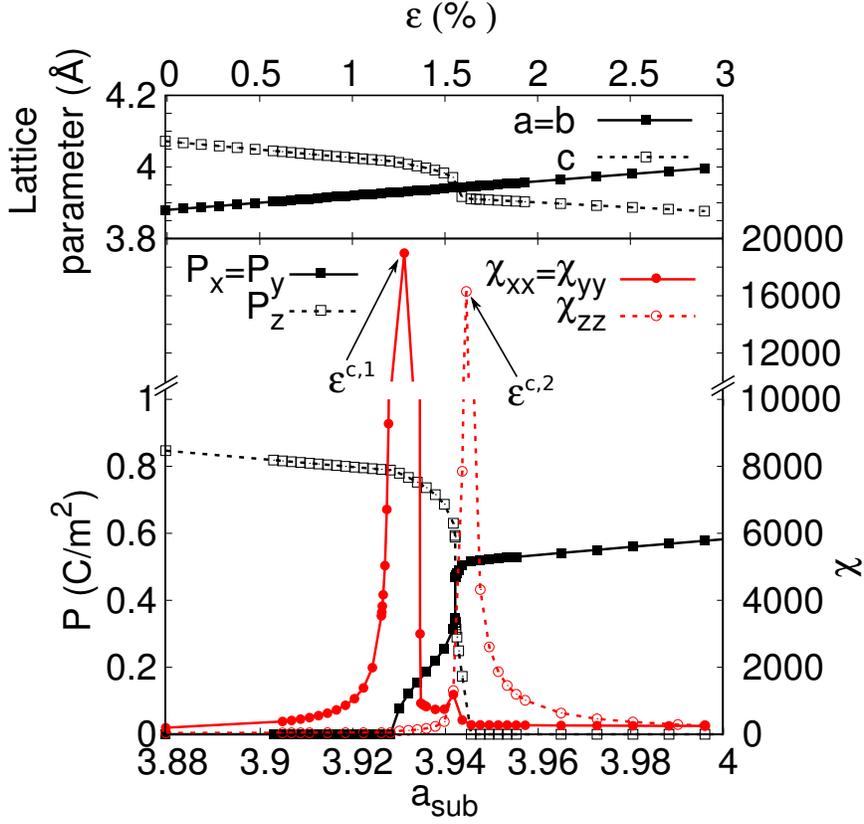}
\caption{(Top panel) Lattice parameters and (main panel) components 
         of the polarization, ${\mathbf P}$ and diagonal terms of the 
         electrical susceptibility tensor, $\bm{\chi}$, as a function 
         of the substrate lattice parameter $a_{sub}$. The
         applied strain referred to the lattice parameter of the 
         zero Kelvin cubic phase is shown in the axis above the 
         top panel. 
        }
\label{fig:chi_vs_e}
\end{figure}

We first investigate the dependence of the polarization on the 
applied strain, i.e. at increasing values of the substrate lattice parameter $a_{sub}$.
As it can be seen in Fig.~\ref{fig:chi_vs_e}, at
a critical strain of $\epsilon^{c,1} \sim 1.25\%$ the polarization ${\mathbf P}$, which was initially 
oriented along the $z$-axis, $P_z \vec{k}$, starts rotating and for 
a strain of $\epsilon^{c,2} \sim 1.73\%$ it lies entirely within the $xy$-plane, 
${\mathbf P}=P_x \vec{i} + P_y \vec{j}$, with $P_x = P_y$.
This behavior has been previously reported in PTO and in other
perovskite oxides and derives from the preferential alignment of
${\mathbf P}$ with the longer axis~\cite{DieguezPRB04,DieguezPRB05,CazorlaPRB15}. Notice that the tetragonal 
phase with space group $P4mm (99)$ is only present when $P_x = P_y = 0$ and 
$P_z \neq 0$. When $P_x = P_y \neq 0$ and 
$P_z \neq 0$, between $\epsilon^{c,1}$ and $\epsilon^{c,2}$,
we find a monoclinic phase $Cm (8)$, and when ${\mathbf P}$ falls into the $xy$-plane
the structure has the orthorhombic symmetry $Amm2 (38)$. The applied biaxial 
tensile strain enlarge the cell in $xy$-plane and shrinks it 
along the $z$-axis (equivalent to the tetragonal axis in the present 
case for $\epsilon = 0$), tending to make the three lattice vectors
similar and thus triggering the rotation of the polarization
(Fig.~\ref{fig:chi_vs_e}, top panel).
Although the reliability of the results obtained within the 
SPDFT approach here employed has been demonstrated in different 
contexts~\cite{ZubkoNature16,ShaferPNAS18,TorresPRM19}, we have explicitly verified that
the ${\mathbf P}(\epsilon)$ curves of Fig.~\ref{fig:chi_vs_e} reproduce
well results obtained within DFT, using the VASP code~\cite{KressePRB96}
with the local density approximation (LDA) and a plane wave cutoff 
of 500~eV with the projector augmented-wave method~\cite{BlochlPRB94,KressePRB99}. 
Note also that, in this work, we are assuming that PTO is in a mono-domain 
state. In reality, multi-domain configurations may form, yielding a 
richer behavior, and experimentally measured values of the thermal 
conductivity are smaller than those estimated here~\cite{TachibanaAPL08}.
Nevertheless, the intrinsic properties of the domains
with rotating polarization should be well captured by our calculations.

The diagonal components of the electrical susceptibility tensor, 
$\bm{\chi}$, are also plotted in Fig.~\ref{fig:chi_vs_e}. When the 
strain-enabled rotation of the polarization is about to occur the 
susceptibility experiences an enormous increase. 
This behavior is an indication that in those conditions a vanishing 
small electric field is sufficient to induce a large change in the 
polarization and the attendant large structural deformation. So, for instance, $\chi_{xx}$,
starts increasing slowly as $a_{sub}$ grows,
but when $\epsilon$ reaches the critical value of 
$\epsilon^{c,1} \sim 1.25\%$ it essentially diverges, a behavior that corresponds to the 
strain-driven onset of $P_x=P_y$ in a second-order transition.  $\chi_{zz}$ has a similar 
behavior when $\epsilon$ is decreased in a highly strained sample 
and reaches $\epsilon^{c,2} \sim 1.73\%$.
In this case we observe a discontinuity in the polarization, with 
the susceptibility reaching very large values as well.
This behavior is typical of weakly first-order transformations, 
which are common in ferroelectric perovskites like PbTiO$_3$~\cite{WojdelPRL14} 
or BaTiO$_3$~\cite{GarciaAPL98}.

The effortless rotation of the polarization close to some critical 
strain values, reflected in the large increase of the susceptibility, 
hints at a similar behavior of the electrophononic coefficients.
Therefore, for each of the strain values considered we have solved
the phonon BTE and computed the thermal conductivity, $\bm{\kappa}$,
first without field and then applying increasingly large field values
along the $x-$ and  $z$-axes to compute the $\bm{\alpha}$ and 
$\bm{\beta}$ tensors. All the results discussed from now on are calculated at 300~K
and before the saturation of the dielectric response.

\begin{figure}[t]
\includegraphics[width=0.7\linewidth]{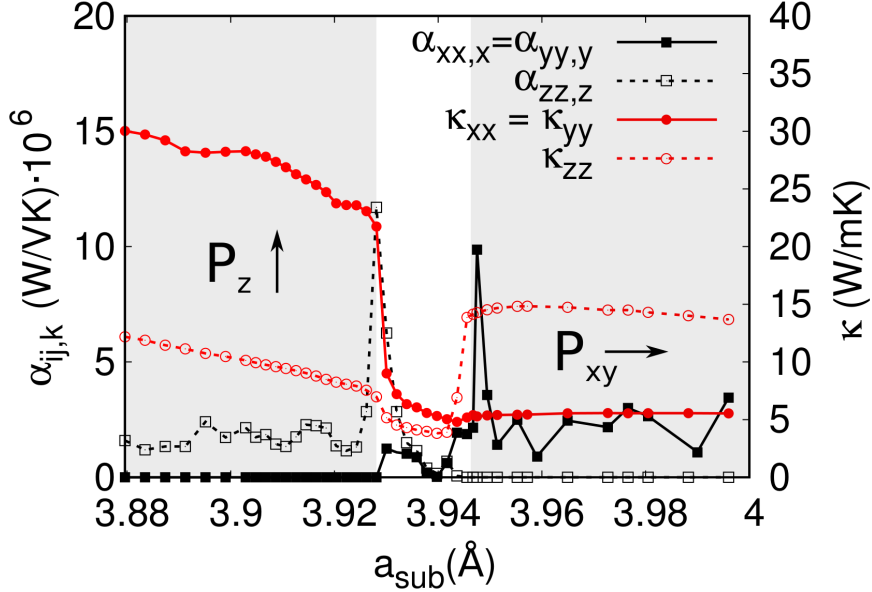}
\caption{Thermal conductivity and first-order electrophononic 
         coefficient, $\alpha_{ii,i}$, as a function of the substrate 
         lattice parameter $a_{sub}$. The polarization direction is 
         represented to indicate when ${\mathbf E}$ is perpendicular or 
         in the plane of ${\mathbf P}$.
         The fluctuation of $\bm{\alpha}$ far from the critical points 
         is caused by the error associated to the polynomial fitting of 
         Eq.~(\ref{eq:response}).
        }
\label{fig:kappa_vs_e}
\end{figure}

In absence of field, as one can observe in Fig.~\ref{fig:kappa_vs_e}, 
the thermal conductivity tensor is anisotropic, as expected in 
a tetragonal or orthorhombic lattice. More precisely, when no strain is applied the 
component parallel to ${\mathbf P}$ ($\kappa_{zz} = 12$~Wm$^{-1}$K$^{-1}$) 
is relatively low and almost three times smaller than the components perpendicular 
to ${\mathbf P}$ ($\kappa_{xx}=\kappa_{yy} = 30$~Wm$^{-1}$K$^{-1}$).
Both $\kappa_{xx}$ and $\kappa_{zz}$ undergo a similar decrease
until the strain-enabled rotation of the polarization takes place.
After that $\kappa_{zz} > \kappa_{xx}$, because now ${\mathbf P}$ lies
in the $xy$-plane, and both components remain roughly constant, with
no significant dependence on strain. The anisotropy of $\bm{\kappa}$
is recovered after the full rotation of ${\mathbf P}$ and the ratio between 
the independent components of $\bm{\kappa}$ is still around 2.5,
though now $\kappa_{ii}$ settle to lower values.
The strain-induced change in the thermal conductivity makes 
$\kappa_{xx}$ and $\kappa_{zz}$ reach a minimum value (4.8 
and 3.8~Wm$^{-1}$K$^{-1}$, respectively) for $\epsilon^{c,1} < \epsilon 
< \epsilon^{c,2}$, when the material is in a monoclinic phase 
with both in-plane and out-of-plane polarization components.
Notice
that by simply applying a biaxial strain to the lattice it is 
possible to achieve a huge reduction of $\bm{\kappa}$. Indeed, 
close to $\epsilon^{c,1}$ and $\epsilon^{c,2}$ , $\kappa_{xx}$
can be reduced a $85~\%$ and $\kappa_{zz}$ a $70~\%$ compared to
their values in the unstrained system.

\begin{figure}[t]
\includegraphics[width=0.7\linewidth]{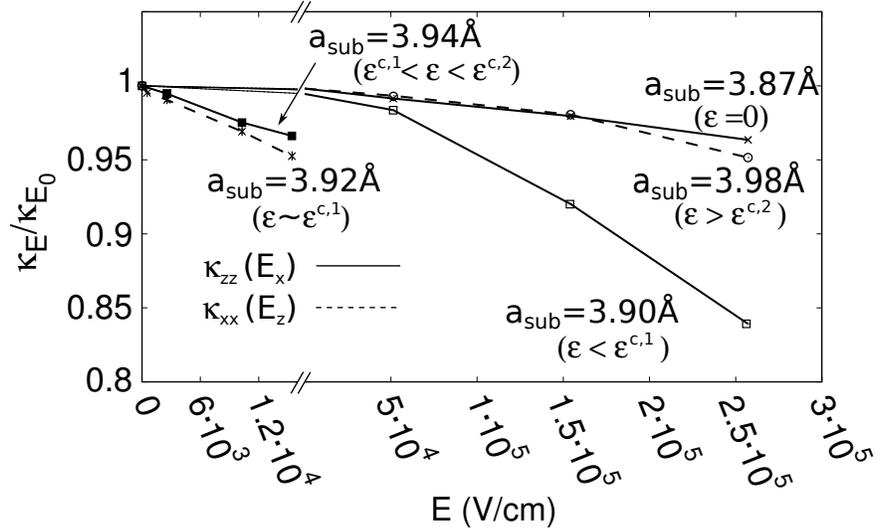}
\caption{Relative change of the thermal conductivity as a function
         of an applied electric field for different values of the 
         biaxial strain. Solid lines show the response in $\kappa_{zz}$
         when a electric field perpendicular to the $z$-axis is applied. Dashed lines
         show the electrophononic response in $\kappa_{xx}$ when the electric field
         is applied parallel to the $z$-axis.
        }
\label{fig:response}
\end{figure}

To study electrophononic effects we have considered fields from $5 \times 10^{4}$ to 
$2.6\times 10^{5}$~V/cm for strain values far from $\epsilon^{c,1}$ and $\epsilon^{c,2}$, 
and smaller values of around $10^{2}$~V/cm, where 
the response of $\bm{\kappa}$ is much stronger (see Fig.~\ref{fig:response}).
The change in $\bm{\kappa}$ as a function of the electric field ${\mathbf E}$ represented
in Fig.~\ref{fig:response} can be fitted to Eq.~(\ref{eq:response}) in order
to obtain the coefficients $\alpha_{ij,k}(T)$ and $\beta_{ij,kl}(T)$
that allow describing the electrophononic response.
By using Eq.~(\ref{eq:response}) we first focus on the linear electrophononic response that relates
$\kappa_{ii}$ to an electric field $E_i$, i.e. $\alpha_{xx,x}=\alpha_{yy,y}$
and $\alpha_{zz,z}$. Our results are shown in Fig.~\ref{fig:kappa_vs_e}.
As it can be seen there, $\alpha_{ii,i}$ has a sizeable value whenever
${\mathbf E}$ is parallel to ${\mathbf P}$, i.e. $\alpha_{zz,z}$ for $\epsilon < 1\%$,
where ${\mathbf P}=P_z \vec{k}$, and $\alpha_{xx,x}=\alpha_{yy,y}$ for $\epsilon > 2\%$,
where ${\mathbf P}=P_x \vec{i} + P_y \vec{j}$. This observation agrees well
with previous reports of a dominant linear electrophononic response for fields parallel to the
tetragonal axis in PTO~\cite{SeijasBellidoPRB18}. Yet, when it approaches the critical
strain values of $\epsilon^{c,1}$ and $\epsilon^{c,2}$, where ${\mathbf P}$ 
spontaneously rotates, $\alpha_{ii,i}$ experiences a sudden and  
large increase. We found an increment of one order of magnitude,
though this gain can in principle be increased by getting closer
to the critical strains.

\begin{figure}[t]
\includegraphics[width=0.7\linewidth]{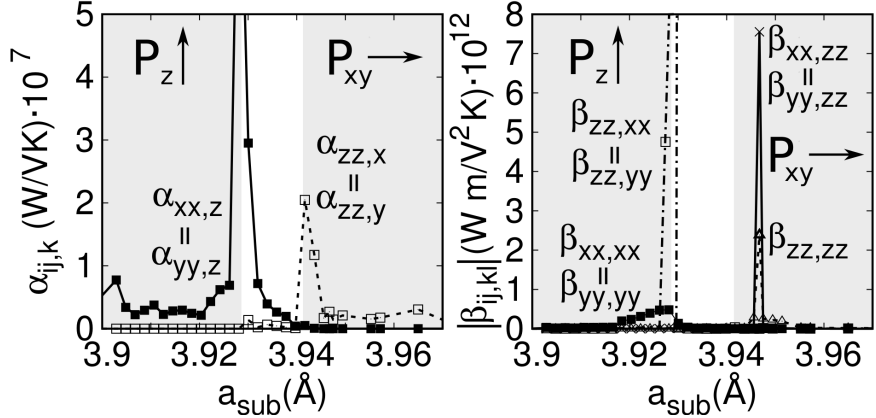}
\caption{(Left) First-order electrophononic coefficient,
         $\alpha_{ii,j}$, with $i \neq j$, and (right) 
         second-order electrophononic coefficient, 
         $\beta_{ii,ii}$ and $\beta_{ii,jj}$, with $i \neq j$, 
         as a function of the substrate lattice parameter $a_{sub}$.
         The polarization direction is
         represented to indicate when ${\mathbf E}$ is perpendicular or
         in the plane of ${\mathbf P}$.
        }
\label{fig:beta-alpha}
\end{figure}

When the applied electric field is perpendicular to the polar axis, 
the electrophononic response of $\bm{\kappa}$ is quadratic, i.e 
$\bm{\beta} \neq 0$, while the linear term $\bm{\alpha}$
is zero by symmetry~\cite{SeijasBellidoPRB18}. This can be observed in
Fig.~\ref{fig:kappa_vs_e} in $\alpha_{xx,x}=\alpha_{yy,y}$ for $\epsilon<\epsilon^{c,1}$, 
where ${\mathbf P}=P_z \vec{k}$, or analogously in $\alpha_{zz,z}$
for $\epsilon>\epsilon^{c,2}$ when ${\mathbf P}=P_x \vec{i} + P_y \vec{j}$.
In Fig.~\ref{fig:beta-alpha} the crossed terms of the linear response $\alpha_{ii,j}$
and $\beta_{ii,jj}$ for $i = j$ and $i \neq j$ are represented. 
From Fig.~\ref{fig:kappa_vs_e} and Fig.~\ref{fig:beta-alpha} we can 
see that when the applied electric field is perpendicular to the 
polarization it always results in a quadratic reduction of the thermal 
conductivity, with negative electrophononic coefficients $\bm{\beta$}, 
being $\bm{\alpha} = 0$. On the contrary, when the applied electric 
field is parallel to the polarization $\bm{\alpha}$ is always positive. 
All the non-zero coefficients tend to become very large close to the critical 
strains.

\begin{figure}[t]
\includegraphics[width=0.7\linewidth]{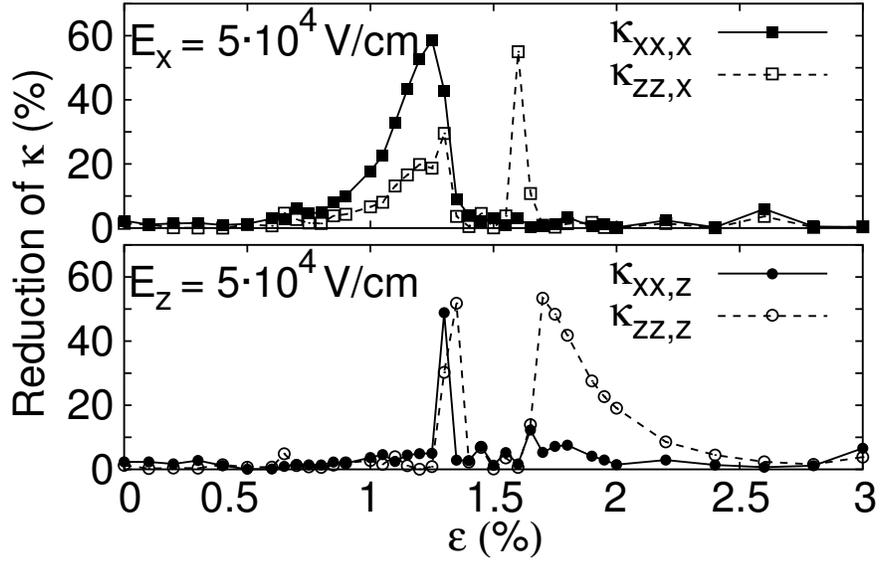}
\caption{Reduction of the lattice thermal conductivity as function 
         of strain for electric fields of $E_i=5\times10^{4}$~V/cm 
         along the x-axis (top) and the z-axis (bottom). The same 
         electrophononic response for $\kappa_{xx,x}$ and $\kappa_{xx,z}$ 
         is found for $\kappa_{yy,y}$ and $\kappa_{yy,z}$.}
\label{fig:reduction}
\end{figure}

To better characterize the electric field induced reduction of the 
thermal conductivity and the interplay of the electrophononic response
with the epitaxial strain, in Fig.~\ref{fig:reduction} we have represented 
the percentage reduction of $\bm{\kappa$} as a function of strain at a 
given electric field. It can be observed that close to the 
critical strains reductions of the thermal conductivity up to $60 \%$
can be easily obtained by applying a small electric field of 
$E_i=5\times10^{4}$~V/cm.
Notice that the overall reduction of the thermal conductivity has two
components: at first it decreases as a result of an applied strain 
close to the critical value, then it further
decreases when the electric field is applied.
These components can be dynamically tuned with high precision by using a piezoelectric
substrate for the former and by an external bias for the latter.

In conclusion we have studied the interplay between the biaxial 
strain and the electrophononic response in a paradigmatic 
ferroelectric perovskite. We have analyzed the tendency to move
from an out-of-plane to an in-plane polarization at a critical
tensile strain and shown that this effortless and spontaneous 
reorientation of the polarization results in giant electrophononic 
responses, at least one order of magnitude larger that those reported in the 
unstrained system, thus enabling the manipulation of the 
phonon flux with vanishingly small electric fields.
Since the underlying physical mechanisms are robust and not specific 
to certain materials or vibrational characteristics, these effects 
can potentially be exploited in a broader class of materials.


\begin{acknowledgments}
We acknowledge financial support by the Ministerio de Econom\'ia,
Industria y Competitividad (MINECO) under grants FEDER-MAT2017-90024-P 
and the Severo Ochoa Centres of Excellence Program
under Grant SEV-2015-0496 and by the Generalitat de Catalunya under grant
no. 2017 SGR 1506.
Work in Luxembourg
was funded by the Luxembourg National Research Fund
(Grant No. FNR/C18/MS/12705883/REFOX/Gonzalez).
We thank the Centro de Supercomputaci\'on de Galicia (CESGA) for the use of
their computational resources.
\end{acknowledgments}




\end{document}